\documentclass[sigconf]{acmart}
\usepackage{enumitem}  % itemize[leftmargin=*]
\usepackage{multirow}  % \multirow
\usepackage{bm}        % \bm
\usepackage{colortbl}  % \cellcolor

\definecolor{tablegray}{HTML}{EFEFEF}

\AtBeginDocument{%
  }

\copyrightyear{2026}
\acmYear{2026}
\setcopyright{cc}
\setcctype{by}
\acmConference[RecSys '26]{20th ACM Conference on Recommender Systems}{September 27-October 02, 2026}{Minneapolis, MN, USA}
\acmBooktitle{20th ACM Conference on Recommender Systems (RecSys '26), September 27-October 02, 2026, Minneapolis, MN, USA}
\acmDOI{10.1145/3773078.3831770}
\acmISBN{979-8-4007-2284-4/2026/09}

\begin{document}

%%
%% The "title" command has an optional parameter,
%% allowing the author to define a "short title" to be used in page headers.
\title{Distribution-Level Contrastive Supervision for Generative Recommendation}

%%
%% The "author" command and its associated commands are used to define
%% the authors and their affiliations.
%% Of note is the shared affiliation of the first two authors, and the
%% "authornote" and "authornotemark" commands
%% used to denote shared contribution to the research.

% Example
% \author{Ben Trovato}
% \authornote{Both authors contributed equally to this research.}
% \email{trovato@corporation.com}
% \orcid{1234-5678-9012}
% \author{G.K.M. Tobin}
% \correspondingauthor
% \authornotemark[1]
% \email{webmaster@marysville-ohio.com}
% \affiliation{%
%   \institution{Institute for Clarity in Documentation}
%   \city{Dublin}
%   \state{Ohio}
%   \country{USA}
% }

% \author{Julius P. Kumquat}
% \correspondingauthor
% \affiliation{%
%   \institution{The Kumquat Consortium}
%   \city{New York}
%   \country{USA}}
% \email{jpkumquat@consortium.net}

\author{Ziqi Xue}
\orcid{0009-0009-0039-8868}
\authornote{Both authors contributed equally to this work.}
\affiliation{
  \institution{University of Science and Technology of China}
  \city{Hefei}
  \country{China}
}
\email{ziqi@mail.ustc.edu.cn}

\author{Dingxian Wang}
\orcid{0000-0002-6880-7869}
\authornotemark[1]
\affiliation{
  \institution{Upwork}
  \city{Seattle}
  \country{USA}
}
\email{dingxianwang@upwork.com}

\author{Yimeng Bai}
\orcid{0009-0008-8874-9409}
\affiliation{
  \institution{University of Science and Technology of China}
  \city{Hefei}
  \country{China}
}
\email{baiyimeng@mail.ustc.edu.cn}

\author{Shuai Zhu}
\orcid{0009-0006-9956-1861}
\affiliation{
  \institution{University of Science and Technology of China}
  \city{Hefei}
  \country{China}
}
\email{zhushuaiozj@mail.ustc.edu.cn}

\author{Jialei Li}
\orcid{0009-0007-9845-0251}
\affiliation{
  \institution{University of Science and Technology of China}
  \city{Hefei}
  \country{China}
}
\email{lijialei.cn@gmail.com}

\author{Xiaoyan Zhao}
\orcid{0000-0001-6001-1260}
% \correspondingauthor
\affiliation{
  \institution{The Chinese University of Hong Kong}
  \city{Hong Kong}
  \country{China}
}
\email{xzhao@se.cuhk.edu.hk}
\authornote{Corresponding author.}

\author{Frank Yang}
\orcid{0009-0001-8117-8788}
\affiliation{
  \institution{Upwork}
  \city{Seattle}
  \country{USA}
}
\email{frankyang@upwork.com}

\author{Andrew Rabinovich}
\orcid{0000-0003-3078-6705}
\affiliation{
  \institution{Upwork}
  \city{Seattle}
  \country{USA}
}
\email{andrewrabinovich@upwork.com}

\author{Yang Zhang}
\orcid{0000-0002-7863-5183}
\affiliation{
  \institution{National University of Singapore}
  \city{Singapore}
  \country{Singapore}
}
\email{zyang1580@gmail.com}

\author{Pablo N. Mendes}
\orcid{0000-0002-0079-7991}
\affiliation{
  \institution{Upwork}
  \city{Seattle}
  \country{USA}
}
\email{Pablomendes@upwork.com}

%%
%% By default, the full list of authors will be used in the page
%% headers. Often, this list is too long, and will overlap
%% other information printed in the page headers. This command allows
%% the author to define a more concise list
%% of authors' names for this purpose.
\renewcommand{\shortauthors}{Xue et al.}

%%
%% The abstract is a short summary of the work to be presented in the
%% article.
\begin{abstract}
% 生成式推荐通过在紧凑的令牌空间中运作，已成为传统检索-排序管道的一种可扩展替代方案。然而，现有方法主要依赖于离散的编码级监督，这会导致信息丢失，并限制了令牌化器与生成式推荐器之间的联合优化。在本研究中，我们提出了一种分布级监督范式，该范式利用多层代码本上的概率分布作为柔性且信息丰富的表示。基于这一思路，我们提出了语义导向分布对齐（SODA），这是一个基于贝叶斯个性化排序的即插即用式对比监督框架，它通过负KL散度对齐语义丰富的分布，同时支持端到端可微分训练。针对多个真实世界数据集的广泛实验表明，SODA 能持续提升各类生成式推荐网络主干的性能，验证了其有效性和普适性。代码将在论文被录用后发布。

% Generative recommendation has emerged as a scalable alternative to traditional retrieve-and-rank pipelines by operating in a compact token space. However, existing methods mainly rely on discrete code-level supervision, which leads to information loss and limits the joint optimization between the tokenizer and the generative recommender. In this work, we propose a distribution-level supervision paradigm that leverages probability distributions over multi-layer codebooks as soft and information-rich representations. Building on this idea, we introduce Semantic-Oriented Distributional Alignment (SODA), a plug-and-play contrastive supervision framework based on Bayesian Personalized Ranking, \xzq{which uses tokenizer-induced codebook distributions as auxiliary supervision targets to align recommender-side representations under a contrastive objective%, while leaving the autoregressive decoding procedure unchanged at inference time
% .} Extensive experiments on multiple real-world datasets demonstrate that SODA consistently improves the performance of various generative recommender backbones, validating its effectiveness and generality. Codes will be available upon acceptance.

Recent generative recommenders improve scalability by retrieving items through token generation instead of traditional ranking over large candidate sets. Yet their training signals are still dominated by discrete code prediction, which overlooks the soft assignment information naturally produced by the tokenizer. This mismatch limits semantic transfer from the tokenizer to the recommender and may hurt overall optimization. We tackle this limitation by introducing a distribution-based supervision scheme for generative recommendation, where multi-level codebook probabilities are treated as soft semantic targets. On top of this design, we develop SODA, a plug-and-play alignment framework that adopts a BPR-style contrastive objective to align recommender representations with target-side distributional representations against negative ones. The proposed method enriches training with finer semantic cues while leaving the decoding stage unchanged. Experimental studies on multiple real-world benchmarks demonstrate that SODA consistently strengthens diverse generative recommendation architectures. Code is available at \url{https://github.com/freyasa/SODA}
\end{abstract}

%%
%% The code below is generated by the tool at http://dl.acm.org/ccs.cfm.
%% Please copy and paste the code instead of the example below.
%%


%%
%% Keywords. The author(s) should pick words that accurately describe
%% the work being presented. Separate the keywords with commas.
%% A "teaser" image appears between the author and affiliation
%% information and the body of the document, and typically spans the
%% page.

%%
%% This command processes the author and affiliation and title
%% information and builds the first part of the formatted document.
\maketitle

\section{Introduction}
\begin{figure*}
    \centering
    \includegraphics[width=0.80\textwidth]{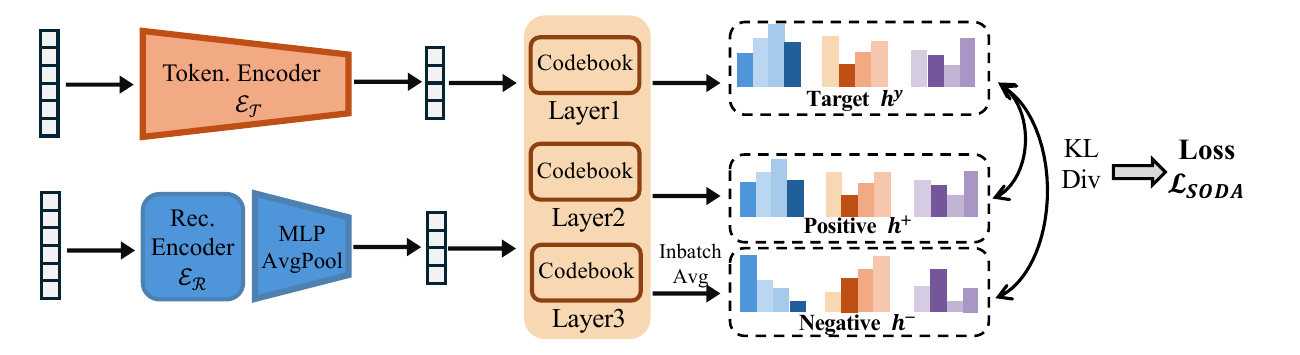}
    \caption{Overview of the proposed SODA framework. SODA leverages probability distributions over the tokenizer’s multi-layer codebooks as soft semantic representations for auxiliary supervision, enabling fine-grained semantic alignment and joint optimization between the tokenizer and the recommender.}
    \label{fig:method}
\end{figure*}

Generative recommendation has recently emerged as a promising paradigm~\cite{GeneRec,hou2025generative,kuaishou_survey,li-etal-2024-large}, attracting growing interest from both academia and industry~\cite{TIGER,Onerec-v2,OnePiece,MTGR}. Typically, it consists of two core components: a tokenizer, which maps each item into a sequence of discrete codes drawn from multi-layer predefined codebooks~\cite{TIGER,LETTER,LC-Rec}, and a generative recommender, which autoregressively predicts the next item based on a tokenized user interaction history. By operating in this compact token space, generative recommenders avoid exhaustive candidate comparisons, enabling scalable and efficient recommendation over large item corpora~\cite{TIGER}, and offering a compelling alternative to traditional retrieve-and-rank pipelines~\cite{Trinity,RankMixer,MERGE}.

Early approaches adopt a decoupled training pipeline, where the tokenizer is first optimized to generate item codes and then frozen during recommender training~\cite{EAGER, SEATER, TIGER, LETTER, DiscRec, LC-Rec}. While straightforward, this separation limits the joint evolution of the two components. To mitigate this, recent work has proposed alternating optimization frameworks, in which the tokenizer and recommender are trained iteratively~\cite{ETEGRec, BLOGER}. From the tokenizer’s perspective, these methods better align code generation with the recommendation objective, enhancing code adaptability. However, even under such stronger coupling, the recommender is still mainly supervised by hard discrete token labels. As a result, the richer distributional assignment information retained by the tokenizer is not explicitly conveyed to the recommender, which may lead to information loss and suboptimal performance~\cite{UniGRec}.

To fully exploit the rich semantic information encoded by the tokenizer, we move beyond code-level supervision and introduce distribution-level supervision. Instead of treating discrete codes as the sole supervisory signal, we leverage the probability distributions over multi-layer codebooks as soft, information-rich representations. Incorporating these distributions as auxiliary supervision allows the recommender to capture fine-grained semantic nuances and improve representation fidelity, which facilitates joint optimization and strengthens the synergy between the two components.

Building on this idea, we propose a novel contrastive supervision strategy based on Bayesian Personalized Ranking (BPR) loss~\cite{BPR} for the recommender, which encourages the model to assign higher scores to positive terms than to negative ones. To extend this principle to distribution-level supervision, the recommender extracts user history representations, which are then mapped to codebook distributions by the tokenizer.

The matching score between a pair of distributions is defined as the negative KL divergence. We term this plug-and-play approach \textit{Semantic-Oriented Distributional Alignment (SODA)}, which encourages the recommender to align semantically rich distributions while enabling effective joint training with the tokenizer. Extensive experiments on multiple real-world datasets show that incorporating SODA into existing generative recommender backbones consistently yields performance gains.

The main contributions of this work are summarized as follows:
\begin{itemize}[leftmargin=*]
\item We introduce distribution-level contrastive supervision for generative recommendation, which enables the recommender to capture fine-grained semantic information encoded by the tokenizer, going beyond traditional code-level signals.

\item We propose SODA, a plug-and-play distribution-level contrastive supervision framework that leverages tokenizer-induced codebook distributions as auxiliary supervision targets for recommender-side semantic alignment.

\item We conduct extensive experiments on multiple real-world datasets,
demonstrating the effectiveness of SODA on multiple generative recommender backbones.
\end{itemize}

% \end{itemize}

% \input{3_pre}

\section{Methodology}

In this section, we first formulate the generative recommendation task, and then provide a presentation of our proposed method.

\subsection{Task Formulation}
We formulate generative recommendation as a sequence-to-sequence modeling task. 
Given a user's interaction history $x$, the objective is to predict the target item $y$.
The framework consists of two coupled components: a tokenizer $\mathcal{T}$ and a generative recommender $\mathcal{R}$.

\vspace{+3pt}
\textbf{Item Tokenization.}
We adopt RQ-VAE~\cite{RQ-VAE} as the tokenizer $\mathcal{T}$ to map each item $i$ with embedding $\bm{z}_i$ into a sequence of discrete tokens. 
The encoder $\mathcal{E}_{\mathcal{T}}$ projects $\bm{z}_i$ into a latent vector, which is recursively quantized with $L$ codebooks $\mathcal{C} = \{\mathcal{C}_l\}_{l=1}^L$ to produce token sequence $(c_1, \dots, c_L)$, and then reconstructed via decoder $\mathcal{D}_{\mathcal{T}}$ to obtain $\tilde{\bm{z}}_i$:
\begin{align}
% \bm{r} &= \mathcal{E}_{\mathcal{T}}(\bm{z}_i), \label{eq:z_i} \\
c_l &= \arg\min_k \lVert \bm{v}_l - \bm{e}_l^k \rVert^2, \quad
\bm{v}_l = \bm{v}_{l-1} - \bm{e}_{l-1}^{c_{l-1}}, \quad \bm{v}_1 = \mathcal{E}_{\mathcal{T}}(\bm{z}_i), \label{eq:quant} \\ 
\tilde{\bm{z}}_i &= \mathcal{D}_{\mathcal{T}}(\sum_{l=1}^L \bm{e}_l^{c_l}),
\end{align}
where $\bm{e}_l^k$ denotes the $k$-th codeword in codebook $\mathcal{C}_l$. The tokenizer is trained to minimize the reconstruction and quantization loss:
\begin{equation}\label{eq:rqvae_loss}
\mathcal{L}_\text{token} =
\lVert \tilde{\bm{z}}_i - \bm{z}_i \rVert^2
+ \sum_{l=1}^{L} \Big(
\lVert \text{sg}(\bm{v}_l) - \bm{e}_l^{c_l} \rVert^2
+ \alpha \lVert \bm{v}_l - \text{sg}(\bm{e}_l^{c_l}) \rVert^2
\Big),
\end{equation}
where $\text{sg}(\cdot)$ denotes the stop-gradient operator and $\alpha$ is a weight.

\vspace{+3pt}
\textbf{Generative Recommendation.}
Given tokenized user history $X$ and target item sequence $Y$ by the tokenizer, the recommender $\mathcal{R}$ is a Transformer encoder--decoder model~\cite{T5} that autoregressively predicts $Y$ from $X$ by encoding $X$ into contextual representations and generating decoder states, trained to minimize the loss:
\begin{equation}\label{eq:gen_loss}
    \mathcal{L}_{\text{rec}} = -\sum_{l=1}^{L} \log P(Y_l \mid X, Y_{<l}).
\end{equation}

\subsection{SODA Framework}
We propose SODA, a plug-and-play framework for distribution-level contrastive supervision beyond discrete token matching, as illustrated in Figure~\ref{fig:method}.

\subsubsection{Supervision Strategy}

Recommendation is inherently a relative matching problem: the current user history should be closer to the target item than to non-matching items or histories. We introduce an additional optimization objective for the recommender based on the Bayesian Personalized Ranking (BPR) loss~\cite{BPR}, which encourages the model to assign higher matching scores to positive representations than to negative ones:
\begin{equation}
    \label{eq:soda_loss}
    \mathcal{L}_{\text{SODA}} = - \log \sigma \left( \beta \cdot (s(h^+, h^y) - s(h^-, h^y)) \right),
\end{equation}
where $h^y$ denotes the target-item distributional representation, and $h^+$ and $h^-$ denote the positive- and negative-history distributional representations, respectively; $h^+$ is derived from the current user, whereas $h^-$ is constructed from other users in the mini-batch, following the common practice of in-batch negative sampling in retrieval and recommendation~\cite{CBNS,XIR,NegSamplingReview}. The distribution matching function $s(\cdot,\cdot)$ is defined based on the negative symmetric Kullback--Leibler (KL) divergence:
\begin{equation}
s(h^a, h^b) = - \Big(\mathrm{KL}(h^a \,\|\, h^b) + \mathrm{KL}(h^b \,\|\, h^a)\Big) / 2.
\end{equation}
\subsubsection{Distributional Representation}
We next present the detailed construction and computation of these representations.

\textbf{Computation of $h^y$.} 
For the target representation $h^y$, we encode its semantic embedding $\bm{z}_y$, measure the resulting latent representation's affinity to all codewords at each layer, and normalize the affinities into probability distributions:
\begin{equation}
    p_{l,k} = \frac{\exp(-\|\bm{v}_l - \bm{e}_l^k\|^2 / \tau)}{\sum_{j=1}^K \exp(-\|\bm{v}_l - \bm{e}_l^j\|^2 / \tau)},
\end{equation}
where $\tau$ is a temperature parameter. 
This formulation serves as a soft relaxation of the hard assignment in Equation~\eqref{eq:quant}, yielding the distributional representation $h^y = \{ p_{l,k}(y) \}_{l=1:L,\; k=1:K}$.

\vspace{+3pt}
\textbf{Computation of $h^+/h^-$.}
The distributional representations $h^+$ and $h^-$ are derived from user interaction histories in the shared tokenizer semantic space. 
Given a user interaction sequence $x$, we encode its tokenized history $X$ with the recommender and project the resulting representation into the tokenizer codebook space:
\begin{equation}
    \bm{r}_x = \text{MLP}(\text{AvgPool}\big( \mathcal{E}_{\mathcal{R}}(X) \big)),
\end{equation}

where $\mathcal{E}_{\mathcal{R}}$ is the recommender encoder. Applying the same soft quantization to $\bm r_x$ yields $h^+=\{p_{l,k}^+\}_{l,k}$, while $h^-=\{p_{l,k}^-\}_{l,k}$ is the layer-wise average of other in-batch users' distributions, providing a stable non-target reference less affected by semantically related false negatives.

\subsection{Training Pipeline}
Following the alternating optimization strategy adopted in prior work~\cite{ETEGRec}, we first pre-train the tokenizer $\mathcal{T}$ to establish a stable discrete latent space using the loss $\mathcal{L}_{\text{token}}$ in Equation~\eqref{eq:rqvae_loss}. 
We then alternately optimize the recommender $\mathcal{R}$ and the tokenizer $\mathcal{T}$. 
Notably, the proposed SODA objective $\mathcal{L}_{\text{SODA}}$ is incorporated into the optimization of the recommender $\mathcal{R}$ during the alternating training process. 
The overall training objective for the recommender is formulated as:
\begin{equation}
    \mathcal{L} = \mathcal{L}_{\text{rec}} + \lambda \, \mathcal{L}_{\text{SODA}},
\end{equation}
where $\lambda$ is a hyperparameter controlling the contribution of the SODA supervision. The overall training objective may additionally include other auxiliary losses when applying SODA to specific generative recommender backbones, such as ETEGRec~\cite{ETEGRec}. SODA incurs no inference overhead and adds less than 5\% training time.

\section{Experiment}
\begin{table}[]
\caption{Statistical details of the evaluation datasets, where “AvgLen” is the average length of historical sequences. }
\label{table:datasets}
\begin{tabular}{cccccc}
\hline
Dataset       & \#User & \#Item & \#Interaction & Sparsity & AvgLen \\ \hline
Beauty        & 22362  & 12083  & 198313        & 99.93\%  & 8.87 \\
Pet           & 19855  & 8498   & 157747        & 99.91\%  & 7.95 \\
Upwork        & 15542  & 33508  & 139217        & 99.97\%  & 8.96 \\ \hline
\end{tabular}
\end{table}

In this section, we conduct experiments to answer the following research questions:

\noindent\textbf{RQ1}: How does SODA perform when applied to existing generative recommendation methods?

\noindent\textbf{RQ2}: What is the contribution of each individual component to the overall effectiveness of SODA?

\begin{table*}[t]
\caption{The overall performance comparisons between the baselines and our method. Results highlighted in bold indicate improvements over the corresponding applied backbone models. {SODA-T}, SODA-L, and {SODA-E} denote applying SODA to TIGER, LETTER, and ETEGRec, respectively.}
\label{table:main_result}
\resizebox{\textwidth}{!}{%
\begin{tabular}{lcccccclcccccc}
\hline
\multicolumn{1}{c}{}                          &                          & \multicolumn{5}{c}{Traditional Method}        &  & \multicolumn{6}{c}{Generative Method}                                                                                                                    \\ \cline{3-7} \cline{9-14} 
\multicolumn{1}{c}{\multirow{-2}{*}{Dataset}} & \multirow{-2}{*}{Metric} & Caser  & GRU4Rec & SASRec & BERT4Rec & HGN    &  & TIGER  & \cellcolor[HTML]{EFEFEF}SODA-T           & LETTER & \cellcolor[HTML]{EFEFEF}SODA-L           & ETEGRec & \cellcolor[HTML]{EFEFEF}SODA-E           \\ \hline
                                              & Recall@10                & 0.0415 & 0.0601  & 0.0550 & 0.0478   & 0.0569 &  & 0.0594 & \cellcolor[HTML]{EFEFEF}\textbf{0.0698} & 0.0670 & \cellcolor[HTML]{EFEFEF}\textbf{0.0741} & 0.0775   & \cellcolor[HTML]{EFEFEF}\textbf{0.0809} \\
                                              & Recall@20                & 0.0639 & 0.0885  & 0.0846 & 0.0735   & 0.0900 &  & 0.0896 & \cellcolor[HTML]{EFEFEF}\textbf{0.1010} & 0.1003 & \cellcolor[HTML]{EFEFEF}\textbf{0.1046} & 0.1149   & \cellcolor[HTML]{EFEFEF}0.1147          \\
                                              & NDCG@10                  & 0.0208 & 0.0336  & 0.0244 & 0.0270   & 0.0277 &  & 0.0320 & \cellcolor[HTML]{EFEFEF}\textbf{0.0385} & 0.0359 & \cellcolor[HTML]{EFEFEF}\textbf{0.0414} & 0.0426   & \cellcolor[HTML]{EFEFEF}\textbf{0.0440} \\
\multirow{-4}{*}{Beauty}                      & NDCG@20                  & 0.0264 & 0.0407  & 0.0318 & 0.0334   & 0.0360 &  & 0.0396 & \cellcolor[HTML]{EFEFEF}\textbf{0.0463} & 0.0443 & \cellcolor[HTML]{EFEFEF}\textbf{0.0491} & 0.0520   & \cellcolor[HTML]{EFEFEF}\textbf{0.0525} \\ \hline
\multicolumn{1}{c}{}                          & Recall@10                & 0.0369 & 0.0504  & 0.0425 & 0.0337   & 0.0519 &  & 0.0541 & \cellcolor[HTML]{EFEFEF}\textbf{0.0672} & 0.0628 & \cellcolor[HTML]{EFEFEF}\textbf{0.0682} & 0.0686   & \cellcolor[HTML]{EFEFEF}\textbf{0.0696} \\
\multicolumn{1}{c}{}                          & Recall@20                & 0.0607 & 0.0771  & 0.0709 & 0.0546   & 0.0843 &  & 0.0860 & \cellcolor[HTML]{EFEFEF}\textbf{0.1000} & 0.0961 & \cellcolor[HTML]{EFEFEF}\textbf{0.1024} & 0.1037   & \cellcolor[HTML]{EFEFEF}\textbf{0.1053} \\
\multicolumn{1}{c}{}                          & NDCG@10                  & 0.0183 & 0.0254  & 0.0202 & 0.0169   & 0.0255 &  & 0.0278 & \cellcolor[HTML]{EFEFEF}\textbf{0.0357} & 0.0323 & \cellcolor[HTML]{EFEFEF}\textbf{0.0357} & 0.0365   & \cellcolor[HTML]{EFEFEF}\textbf{0.0369} \\
\multicolumn{1}{c}{\multirow{-4}{*}{Pet}}     & NDCG@20                  & 0.0243 & 0.0321  & 0.0273 & 0.0221   & 0.0337 &  & 0.0358 & \cellcolor[HTML]{EFEFEF}\textbf{0.0440} & 0.0407 & \cellcolor[HTML]{EFEFEF}\textbf{0.0443} & 0.0453   & \cellcolor[HTML]{EFEFEF}\textbf{0.0459} \\ \hline
                                              & Recall@10                & 0.0455 & 0.0537  & 0.0872 & 0.0321   & 0.0516 &  & 0.0885 & \cellcolor[HTML]{EFEFEF}\textbf{0.1012} & 0.0730 & \cellcolor[HTML]{EFEFEF}\textbf{0.0811} & 0.0934   & \cellcolor[HTML]{EFEFEF}\textbf{0.0935}          \\
                                              & Recall@20                & 0.0730 & 0.0855  & 0.1264 & 0.0516   & 0.0812 &  & 0.1380 & \cellcolor[HTML]{EFEFEF}\textbf{0.1568} & 0.1202 & \cellcolor[HTML]{EFEFEF}\textbf{0.1257} & 0.1405   & \cellcolor[HTML]{EFEFEF}\textbf{0.1421}          \\
                                              & NDCG@10                  & 0.0231 & 0.0272  & 0.0432 & 0.0161   & 0.0270 &  & 0.0441 & \cellcolor[HTML]{EFEFEF}\textbf{0.0508} & 0.0356 & \cellcolor[HTML]{EFEFEF}\textbf{0.0403} & 0.0465   & \cellcolor[HTML]{EFEFEF}\textbf{0.0467}          \\
\multirow{-4}{*}{Upwork}                      & NDCG@20                  & 0.0300 & 0.0352  & 0.0531 & 0.0210   & 0.0344 &  & 0.0566 & \cellcolor[HTML]{EFEFEF}\textbf{0.0647} & 0.0474 & \cellcolor[HTML]{EFEFEF}\textbf{0.0515} & 0.0583   & \cellcolor[HTML]{EFEFEF}\textbf{0.0589} \\ \hline
\end{tabular}
}
\end{table*}

\subsection{Experimental Setup}

\subsubsection{Datasets}

Our experiments are conducted on three real-world datasets: two publicly available Amazon Review benchmarks (\textbf{Beauty} and \textbf{Pet})\footnote{\url{https://jmcauley.ucsd.edu/data/amazon/index_2014.html}} and a proprietary dataset from our online freelancing platform (\textbf{Upwork}). In Upwork, each interaction is a timestamped hiring event, with employers as users and freelancers as items. Table~\ref{table:datasets} reports the statistics of all datasets. Following prior studies~\cite{TIGER,SASRec}, we apply 5-core filtering, keeping only users and items with at least five interactions. Each user sequence is truncated or padded to length 20 based on the most recent interactions. Evaluation follows a leave-one-out split.

\subsubsection{Baselines}
\label{sec:exp_baselines}

The baseline methods are divided into two categories as follows: (1) \textit{Traditional recommendation methods:} \textbf{Caser}~\cite{Caser}, \textbf{GRU4Rec}~\cite{GRU4Rec}, \textbf{SASRec}~\cite{SASRec}, \textbf{BERT4Rec}~\cite{BERT4Rec}, and \textbf{HGN}~\cite{HGN}; and (2) \textit{Generative recommendation methods:} \textbf{TIGER}~\cite{TIGER}, \textbf{LETTER}~\cite{LETTER}, and \textbf{ETEGRec}~\cite{ETEGRec}.  
Since SODA is a plug-and-play framework, we applied it to each generative backbone, yielding \textbf{SODA-T}, \textbf{SODA-L}, and \textbf{SODA-E}, respectively, for comparison.

\subsubsection{Evaluation Metrics}
\label{sec:exp_metrics}
We adopt two widely used metrics for top-$K$ recommendation: \textbf{Recall@$K$} and \textbf{NDCG@$K$} with $K \in \{10, 20\}$. To ensure an unbiased evaluation, we perform {full ranking} over the entire item set instead of sampling negatives. For all generative methods, we use a {constrained beam search}~\cite{LETTER} with a beam size of 30 during inference.

\subsubsection{Implementation Details}
\label{sec:exp_details}

For our proposed method, we adopt RQ-VAE as the tokenizer model~\cite{RQ-VAE} and T5~\cite{T5} as the recommender model. 
The initial item semantic embeddings are extracted using \texttt{sentence-t5-base}~\cite{ni-etal-2022-sentence} from item titles and descriptions for the Beauty and Pet datasets (768 dimensions). 
We tune the hyperparameter $\lambda$ over $\{1e\!-\!2, \dots, 1e\!-\!6\}$ in orders of magnitude, the temperature $\tau$ over $\{1, \dots, 1e\!-\!5\}$, and the scaling factor $\beta$ over $\{1, \dots, 1000\}$.
For traditional baselines, we use the implementations provided in the open-source library RecBole~\cite{recbole}.
All other generative recommendation model configurations and training settings follow prior work~\cite{ETEGRec}.

\begin{table}[t]
\caption{Results of ablation study of SODA on Beauty.}
\label{table:ablation}
\resizebox{0.475\textwidth}{!}{%
\begin{tabular}{lcccc}
\hline
Method          & Recall@10       & Recall@20       & NDCG@10         & NDCG@20         \\ \hline
\textbf{SODA-T} & \textbf{0.0698} & \textbf{0.1010} & \textbf{0.0385} & \textbf{0.0463} \\
w/o neg    & 0.0639          & 0.0959          & 0.0366          & 0.0447          \\
w/o loss        & 0.0653          & 0.0978          & 0.0367          & 0.0449          \\
w/o alter & 0.0594          & 0.0896          & 0.0320          & 0.0396          \\ \hline
\textbf{SODA-L} & \textbf{0.0741} & \textbf{0.1046} & \textbf{0.0414} & \textbf{0.0491} \\
w/o neg    & 0.0685          & 0.0979          & 0.0387          & 0.0461          \\
w/o loss        & 0.0653          & 0.0969          & 0.0376          & 0.0456          \\
w/o alter & 0.0670          & 0.1003          & 0.0359          & 0.0443          \\ \hline
\end{tabular}
}
\end{table}

\subsection{Main Result (RQ1)}
Table~\ref{table:main_result} presents the overall performance of different models across the three datasets. Based on the analysis of the experimental results, we draw the following key conclusions:
\begin{itemize}[leftmargin=*]
\item \textbf{Applying SODA to various generative recommender backbones consistently enhances their performance.} This superiority can be attributed to the distribution-level supervision, which mitigates the information loss caused by the rigid discretization in code-based methods. By leveraging probability distributions as soft representations, our approach captures fine-grained semantic differences.

% \vspace{-2pt}
\item \textbf{Generative methods generally outperform traditional ID-based baseline methods.} This suggests that generative approaches can capture complex collaborative signals while leveraging rich semantic information from the tokenizer (instead of random IDs), thereby validating the effectiveness of the generative paradigm in recommendation. 

\end{itemize}

\subsection{Ablation Study (RQ2)}

To evaluate the contribution of each key component in SODA, we conduct an ablation study by progressively removing individual components. Due to space limitations, we report results for SODA-T and SODA-L on the Beauty dataset; results on other datasets and backbones show consistent trends. Specifically, we compare the following variants: 
% \begin{itemize}[leftmargin=*]
\begin{itemize}[leftmargin=*,topsep=2pt,itemsep=0pt,parsep=0pt,partopsep=0pt]
\item \textbf{w/o neg}, which replaces the BPR-based contrastive objective with a pointwise loss by removing the negative term and modeling only the positive history--target relation; 
\item \textbf{w/o loss}, which completely removes $\mathcal{L}_{\text{SODA}}$ and relies solely on discrete code-level supervision; 
\item \textbf{w/o alter}, which adopts standard TIGER/LETTER training with decoupled optimization instead of alternating training.
\end{itemize}
From the results in Table~\ref{table:ablation}, we draw the following observations:
% \begin{itemize}[leftmargin=*]
\begin{itemize}[leftmargin=*,topsep=2pt,itemsep=0pt,parsep=0pt,partopsep=0pt]
\item \textbf{Impact of Our Supervision.} 

Both \textbf{w/o neg} and \textbf{w/o loss} consistently degrade performance, indicating that distribution-level supervision provides fine-grained semantic signals beyond code-level supervision. Contrastive negative samples further improve representation discrimination and semantic alignment between user histories and target items.

\item \textbf{Impact of Alternating Optimization.} 
The \textbf{w/o alter} variant underperforms because removing alternating optimization fixes item codes and prevents their adaptation to recommendation contexts, demonstrating the benefit of dynamically updated, recommendation-aware token representations.

\end{itemize}

\section{Conclusion}

We introduced SODA, a plug-and-play distribution-level supervision framework that aligns recommender representations with tokenizer-induced codebook distributions through a BPR objective based on negative symmetric KL divergence. By exploiting fine-grained semantics beyond discrete codes, SODA consistently improves multiple generative recommender backbones without inference overhead.

For future work, we aim to evaluate the scalability of our method on large-scale industrial recommendation datasets to further assess its robustness and efficiency when applied to extremely large item corpora.

%%
%% The next two lines define the bibliography style to be used, and
%% the bibliography file.
\bibliographystyle{ACM-Reference-Format}
\balance
\bibliography{8_ref}

\end{document}